\documentclass{ws-ijmpd}
\usepackage[super,compress]{cite}
\usepackage{url}           
\usepackage[english]{babel}
\usepackage{xcolor}
\usepackage{changes}
\usepackage{graphicx}

\usepackage{savesym}
\savesymbol{captionbox}
\usepackage{subcaption}
\restoresymbol{TXF}{captionbox}
\usepackage{multirow}
\usepackage{siunitx}
\DeclareSIUnit\parsec{pc}
\DeclareSIUnit\lightyear{ly}
\newcommand{\pa}[1]{\left(#1\right)}	
\newcommand{\pac}[1]{\left[#1\right]}

\font\bigastfont=cmr10 scaled \magstep 1
\newcommand{\bdot}{\hbox{\bigastfont .}}

\newcommand{\CQ}{{\mathcal Q}}

\newcommand{\CX}{{\mathcal X}}
\newcommand{\CL}{{\mathcal L}}

\newcommand{\dd}{{\mathrm d}}

\newcommand{\CD}{\mathcal{D}}

\newcommand{\CR}{\mathcal{R}}

\newcommand{\average}[2]{\left\langle #1 \right\rangle_{\cal #2}}
\newcommand{\laverage}[2]{\left\langle #1 \right\rangle_{\mathcal{#2}_{\rm \bf i}}}

\newcommand{\initial}[1]{{#1_{\rm \bf i}}}
\newcommand{\now}[1]{{#1_{\it 0}}}

\newcommand{\say}[1]{`#1'}
\usepackage{hyperref}
{
 \definecolor{BLACK}{gray}{0}
 \definecolor{WHITE}{gray}{1}
 \definecolor{RED}{rgb}{1,0,0}
 \definecolor{GREEN}{rgb}{0,1,0}
\definecolor{dgreen}{rgb}{.1,.6,.1}
\definecolor{BLUE}{rgb}{0,0,1}
 \definecolor{CYAN}{cmyk}{1,0,0,0}
 \definecolor{MAGENTA}{cmyk}{0,1,0,0}
 \definecolor{YELLOW}{cmyk}{0,0,1,0}
 \definecolor{aw}{rgb}{0.2,0.5,0.75}
  }
  \hypersetup{
    colorlinks,%
    citecolor=blue,%
    filecolor=blue,%
    linkcolor=magenta,%
    urlcolor=blue
}

\definecolor{MyR}{rgb}{0.9,0,0}

\definecolor{MyDarkRed}{rgb}{0.7,0,0}

\definecolor{PineGreen}{rgb}{0.0, 0.47, 0.44}

\begin{document}

\def\PRL{{Phys.\ Rev.\ Lett.}}
\def\JCAP{{\it J.\ Cosmol.\ Astropart.\ Phys.}\ JCAP}
\def\AJ{{Astron.\ J.}}
\def\ApJ{{Astrophys.\ J.}}
\def\ApJL{{Astrophys.\ J.}}
\def\PR{{Phys.\ Rev.}}
\def\MNRAS{{Mon.\ Not.\ R.\ Astr.\ Soc.}}
\def\CQG{{Class.\ Quantum Grav.}}
\def\GRG{{Gen.\ Relativ.\ Grav.}}
\def\IJMP{{Int.\ J.\ Mod.\ Phys.}}
\def\AA{{Astron. \& Astrophys.}}
\def\PRD{{{Phys.\ Rev. D.}}}

\markboth{C\'ELIA DESGRANGE, ASTA HEINESEN AND THOMAS BUCHERT}
{ \uppercase{Dynamical spatial curvature as a fit to type Ia supernovae}}

%
\catchline{}{}{}{}{}
%

\title{ \uppercase{Dynamical spatial curvature as a fit to type Ia supernovae}}

\author{C\'ELIA DESGRANGE$^1$, ASTA HEINESEN$^2$ AND THOMAS BUCHERT$^1$}

\address{$^1$Univ Lyon, ENS de Lyon, Univ Lyon1, CNRS, \\Centre de Recherche Astrophysique de Lyon UMR5574 \\ Lyon, F--69007, France \\
Emails: celia.desgrange@ens--lyon.fr $\ $buchert@ens--lyon.fr }

\address{$^2$School of Physical \& Chemical Sciences, University of Canterbury,
Private Bag 4800 \\ Christchurch 8140, New Zealand \\ 
Email: asta.heinesen@pg.canterbury.ac.nz}

\maketitle


\begin{abstract}
Few statements in cosmology can be made without assuming a cosmological model within which to interpret data. 
Statements about cosmic acceleration {are} no exception to this rule, and the inferred positive volume acceleration of our Universe often quoted in the literature is valid in the context of the standard Friedmann-Lema\^{\i}tre-Robertson-Walker (FLRW) class of space-times. 

Using the Joint Light-curve Analysis (JLA) catalogue of type Ia supernovae (SNIa), we examine the fit of a class of exact scaling solutions with dynamical spatial curvature formulated in the framework of a scalar averaging scheme for relativistic inhomogeneous space-times. 
In these models, global volume acceleration may emerge as a result of the non-local variance between expansion rates of clusters and voids, the latter gaining volume dominance in the late-epoch Universe.

We find best-fit parameters for a scaling model of backreaction that are reasonably consistent with previously found constraints from SNIa, CMB, and baryon acoustic oscillations data.
The quality of fit of the scaling solutions is indistinguishable from that of the $\Lambda$CDM model and the timescape cosmology from an Akaike Information Criterion (AIC) perspective. 
This indicates that a broad class of models can account for the $z\lesssim 1$ expansion history.
\end{abstract}


\ccode{PACS numbers: 98.80.-k, 98.80.Es, 98.80.Jk, 95.36.+x}

%
%

\section{Introduction \label{sec:Intro}}
Supernovae type Ia, and their use as approximate standard candles, have led to one of the most remarkable observations made in cosmology: the volume growth of the Universe, when interpreted within the Friedmann-Lema\^{\i}tre-Robertson-Walker (FLRW) class of models, is accelerating \cite{acceleration_discovery_Riess,acceleration_discovery_Perlmutter,acceleration_perlmutter_1998}.    

This discovery, together with BAO (baryon acoustic oscillations) features in the galaxy distribution and the CMB (cosmic microwave background), serves as a further observational cornerstone of the $\Lambda$CDM paradigm (cold dark matter with dark energy modeled by a positive cosmological constant $\Lambda$) as the most successful expansion history within the FLRW class of models.
The largely self-consistent fit of the $\Lambda$CDM model to data comes at the expense of introducing unknown energy-momentum sources. There are also a number of observational tensions \cite{WendyFreedman,clustercounts,lymanalpha,primordialLithium,kids,bowman,steinhardt,sheth}; for a recent overview see Ref. \refcite{tensions}.

Given these mysteries encountered when interpreting cosmological data in the FLRW models, it might be worth to reconsider their status as being the almost exclusively studied solutions of the Einstein equations in cosmology. 
The FLRW metrics offer a simple framework within which to interpret observations, but are extremely limited in their dynamical features. The extent to which the standard FLRW models can accurately serve as a global \say{background} and provide an average description for the all-scale hierarchy of structures in the Universe must be assessed. 

The real Universe is not spatially homogeneous and isotropic, but is at best associated with \emph{statistical} symmetries on \say{cosmological scales}.  
Whether a statistical description properly marginalizing over the hierarchy of structures in our Universe
will yield an FLRW space-time as an accurate description of large-scale cosmic dynamics is not evident from first principles. 

In our analysis we employ a covariant scalar averaging scheme \cite{Buchert2000,Buchert2001} appropriate for marginalizing over structure in a general relativistic cosmological fluid description to obtain global evolution equations analogous to Friedmann's equations for space-times with no imposed symmetries.
One of the main insights of this scalar averaging scheme is that average spatial curvature is gravitationally unstable: \cite{roy:instability} maintaining (close to) zero global spatial curvature (which is a built in feature of the $\Lambda$CDM paradigm of cosmology) in a general relativistic universe model with structure requires fine-tuning at all stages of the expansion history.
In the averaging scheme which we employ, the FLRW spatial curvature behaviour does not follow naturally as an averaged description in universe models without assuming \emph{exact} spatial homogeneity and isotropy. 
In particular, the flat $\Lambda$CDM model constrains spatial curvature to remain zero for all times, a property which is in general not recovered when averaging a lumpy space-time.

In this paper we consider a class of \say{scaling solutions} \cite{morphon2006,larena2009,roy:instability}, which forms a closure condition for the system of general cosmological equations for averaged scalar variables.  
These solutions have average spatial curvature evolution which is fundamentally different from that of the FLRW class of space-times.
Some observational tests have already been made with these scaling solutions in Ref.~\refcite{larena2009}, using CMB data and a sparse SNIa sample, and in Ref.~\refcite{Chinese2018}, using BAO measurements and the differential age method. 

Another model built from the same scalar averaging scheme as the scaling solutions, the \say{timescape model},\footnote{For a review of the timescape model see Ref. \refcite{TS2009forma}.} has been tested on the Joint Light-curve Analysis (JLA)\cite{JLA2014} catalogue of type Ia supernovae and showed an equally good fit to that of the spatially flat $\Lambda$CDM model.
The successful fit of the timescape model suggests that spatial curvature evolution has the potential of mimicking dark energy in the late epoch Universe. 
Curvature evolution in the late epoch Universe has first been applied to supernova data by Kasai \cite{kasai} by dividing the supernova sample into early- and late-type 
subsamples and fitting these two subsamples with different FLRW models, treating the respective FLRW curvature parameters as free parameters in the analysis.
While it is known that the FLRW model with negative constant curvature does not successfully fit cosmological data, nothing in this result prevents non-FLRW curvature evolution towards present-epoch negative curvature---as expected from general considerations of averaged inhomogeneous universe models \cite{buchertcarfora}. 

In this work we use the JLA catalogue to test a family of scaling solutions for the average variables entering in the scalar averaging scheme using the Spectral Adaptive Lightcurve Template 2 (SALT2) relation.
We will compare the resulting fit to that of the $\Lambda$CDM model, the empty universe model,\footnote{While the empty universe model is unphysical and ruled out by combined constraints from CMB, SNIa, and BAO data, it is an interesting idealization for the \emph{late-epoch Universe} in which matter is highly clustered within tiny volumes and photons primarily propagate in large, empty void-regions.} and the timescape model. 

In Sec. \ref{sec:formalism} we review the scalar averaging scheme and the scaling solutions employed in this paper, and we provide the distance modulus--redshift relation for the scaling solutions. In Sec. \ref{sec:sn} we briefly describe the SALT2 method for standardising supernovae, and we review the likelihood-function used in the statistical analysis of the JLA catalogue. In Sec. \ref{sec:modelanalysis} we present the results of our analysis: constraints on model parameters of the investigated scaling solutions, and the quality of fit as compared to that of the $\Lambda$CDM model, the empty Milne model (i.e. the FLRW model without sources, but negative constant curvature), and the timescape model.
In Sec. \ref{sec:curvaturedynamics} we examine a FLRW curvature consistency measure, compute the analogous measure for the best-fit results for the scaling solutions, and discuss the potential use of this measure for the discrimination between FLRW models and backreaction models with emerging deviations from the FLRW constant curvature geometry in future surveys. We conclude in Sec. \ref{sec:conclusion}. 

\section{The scalar averaging scheme and scaling solutions} 
\label{sec:formalism}
We now recall the class of scaling solutions of the scalar averaging scheme and provide an associated distance modulus--redshift relation, which we are going to test in this paper. 

We base our analysis on a scheme for averaging of scalar variables in a self-gravitating dust-fluid, recalled in Sec. \ref{subsec:buchertsscheme}, and formulated  in terms of effective cosmological parameters in Sec. \ref{subsec:cosmologicalparams}. 
In Sec. \ref{subsec:scalingsol} we introduce the scaling solutions, and in Sec. \ref{subsec:templatemetric} we describe our procedure for constructing an effective metric, a so-called template metric, to match an effective light cone structure to the large-scale model defined in the averaging scheme. 
From this prescribed metric we finally obtain the expressions for the distance modulus--redshift relation in 
Sec. \ref{subsec:distancemodulus}.

\subsection{Irrotational dust averages} \label{subsec:buchertsscheme}
We consider a Lorentzian manifold with a self-gravitating irrotational dust fluid as the energy-momentum source in the Einstein equations. The aim is to describe average dynamical properties of this system, and to determine an effective description of light propagation on cosmological scales without knowing the metric of the lumpy space-time in detail. 

The exact scalar averaging scheme we employ is a method for obtaining global dynamical equations for such a space-time, without knowledge of its \say{micro state}. 
Here we provide the relevant dynamical equations for this analysis with a short explanation of the relevant variables. 
Precise definitions of the variables and the averaging operation, and the full derivation of the below equations can be found in Ref.~\refcite{Buchert2000}. 
Throughout this paper we work in units of $c = 1$, $c$ being the speed of light in vacuum.

Let $\bm u = - \bm \nabla t$ be the $4-$velocity field of the fluid source, with $t$ being a proper time function of the fluid, and let $\varrho$ be its rest mass density. 
From averaging the local Raychaudhuri equation in the fluid rest frame over a spatial domain $\CD$ comoving with the fluid (no net-flux of particle world-lines through the boundaries of the averaging domain), we obtain the \emph{averaged Raychaudhuri equation},
\begin{equation}
\label{eq:averagedraychaudhuri}
3\,\frac{{\ddot a}_\CD}{a_\CD} \,+\, 4\pi G \, \average{\varrho}{\CD} \,-\,\Lambda\;=\; {\CQ}_\CD\,,
\end{equation}
where $a_\CD$ is the volume scale factor, $\average{.}{\CD}$ denotes {covariant} averaging in the fluid frame over the comoving spatial domain $\CD$, $\Lambda$ is the cosmological constant,\footnote{We set $\Lambda=0$ in the investigations of this paper, as we investigate averaged models without dark energy, but keep $\Lambda$ in the equations of this section for completeness.} and the overdot denotes the covariant time-derivative{: $\dot{} \equiv \frac{\dd}{\dd t}$}.

{Note that in general $\average{S}{\CD}^{\bdot} \neq \langle \dot S \rangle_\CD$, where $\average{S}{\CD}^{\bdot}$ is the time-derivative of the averaged variable $\average{S}{\CD}$, and $\langle \dot S \rangle_\CD$ is the average of the time-derived local scalar $\dot{S} = u^{\mu}\nabla_{\mu}S$. For details on the averaging operation and the non-commutativity of averaging and time-derivative, see Ref.~\refcite{Buchert2000}.}

$\CQ_\CD$ is the \say{kinematical backreaction} which is defined from the variance of the rate of expansion of the fluid congruence and the averaged shear of the fluid congruence over the domain $\CD$.

The local energy constraint equation can be averaged in a similar way to obtain the averaged energy constraint equation,
\begin{equation}
\label{eq:averagedhamilton}
3\left( \frac{{\dot a}_\CD}{a_\CD}\right)^2 \,-\, 8\pi G \,\average{\varrho}{\CD}\,-\,\Lambda \;=\; - \, \frac{\average{\CR}{\CD}\,+\,{\CQ}_\CD }{2}\,,
\end{equation}
where $\average{\CR}{\CD}$ is the averaged spatial scalar curvature. 
Finally, we have the average of the local energy-momentum conservation equation,  
\begin{equation}
\label{eq:energyconservation}
\average{\varrho}{\CD}^{\bdot} \,+\, 3\, \frac{{\dot a}_\CD}{a_\CD}\average{\varrho}{\CD}\;=\;0 \,.
\end{equation}
All of the global variables $a_\CD$, $\average{\varrho}{\CD}$, ${\CQ}_\CD$, and $\average{\CR}{\CD}$ entering in the averaged equations depend on the proper time slice parameterized by $t$ and the spatial domain of integration $\CD$.

Note that when positive, $\CQ_\CD$ can act as an effective source for global acceleration in (\ref{eq:averagedraychaudhuri}). ${\CQ}_\CD$ will in general depend on cosmic time $t$, and on spatial scale through the dependence on the domain of averaging.

Combining (\ref{eq:averagedraychaudhuri}), (\ref{eq:averagedhamilton}), and (\ref{eq:energyconservation}), the variables have to obey the following \textit{integrability condition}:
\begin{equation}
\label{eq:integrability}
\frac{1}{a_\CD^6} \, (\,{\CQ}_\CD \,\;a_\CD^6 \,)^{\bdot}
\,+\, \frac{1}{a_\CD^{2}} \,(\,\average{\CR}{\CD}a_\CD^2 \,)^{\bdot}\, \;=\;0\,,
\end{equation}
{where the notation $(.)^{\bdot}$ means differentiation with respect to $t$ of the entire content of the parenthesis.}
Eq. (\ref{eq:integrability}) shows that kinematical backreaction and the averaged spatial curvature are coupled. 
This equation is key to understanding the evolution of global curvature as a consequence of structure formation.
Note that by demanding ${\CQ}_\CD \propto 1/ a_\CD^6$ (including the trivial case ${\CQ}_\CD = 0$), the averaged curvature obeys a separate (scale-dependent) conservation equation corresponding to the FLRW curvature constraint $(\,\average{\CR}{\CD}a_\CD^2 \,)^{\bdot} = 0$. 

\subsection{Cosmological parameters} \label{subsec:cosmologicalparams}
It shall be convenient to write the averaged energy constraint equation (\ref{eq:averagedhamilton}) in terms of effective cosmological parameters \cite{Buchert2008}.
Dividing (\ref{eq:averagedhamilton}) by $(3 \,H_\CD^2)$, where we call the functional $H_\CD \equiv \dot{a}_\CD/a_\CD$ \say{the global Hubble parameter}, we have:
\begin{equation}
\label{eq:cosmicquartet}
\Omega_m^\CD \,+\, \Omega_\Lambda^\CD \,+\, \Omega_\CR^\CD \,+\, \Omega_\CQ^\CD\;=\;1\ ,
\end{equation}
where the four cosmological \say{parameters} $\Omega_m^\CD , \Omega_\Lambda^\CD , \Omega_\CR^\CD$, and $\Omega_\CQ^\CD$ constitute the \say{cosmic quartet} and are defined by:
\begin{eqnarray}
\label{eq:omega}
\Omega_{m}^{\CD} &\;\equiv\; \displaystyle{\frac{8\pi G}{3 H_{\CD}^2} \, \average{\varrho}{\CD}}\ ; \qquad \Omega_{\Lambda}^{\CD} &\;\equiv\; \;  \frac{\Lambda}{3 H_{\CD}^2} \ ; \\ 
\Omega_\CR^\CD &\;\equiv\;  - \, \displaystyle{\frac{\average{\CR}{\CD}}{6 H_\CD^2}} \ ; \qquad \Omega_{\CQ}^{\CD} &\;\equiv\;  - \,\frac{{\CQ}_{\CD}}{6 H_{\CD}^2 }\ .
\end{eqnarray}
As we wish to see whether the averaged spatial curvature $\Omega_\CR^\CD$ and backreaction $\Omega_\CQ^\CD$ cosmological parameters can mimic dark energy without a local energy component violating the strong energy condition, we set $\Omega_{\Lambda}^{\CD} = 0$. 
We can further rewrite (\ref{eq:cosmicquartet}) in terms of deviations from a spatially flat Friedmannian parametrization,
\begin{equation}
\label{eq:omX}
\Omega_{m}^{\CD} \,+\, \Omega_\CX^{\CD}\;=\;1\ \ ;\qquad \Omega_\CX^\CD \; \equiv \; \Omega_{\CR}^{\CD} \,+\, \Omega_{\CQ}^{\CD} \ , 
\end{equation}
where $\CX$ stands for \say{$\CX-$matter}: an effective \say{matter} cosmological component that has the potential to mimic dark energy and/or dark matter signatures as they appear in the standard $\Lambda$CDM model.

\subsection{Scaling solutions to the averaged Einstein equations} 
\label{subsec:scalingsol}
In order to uniquely determine the solutions to the four unknown functions $a_\CD$, $\average{\varrho}{\CD}$, $\average{\CR}{\CD}$, and ${\CQ}_\CD$ satisfying the equations (\ref{eq:averagedraychaudhuri})--(\ref{eq:integrability}) (where one of the equations in the set is redundant), we must specify one additional equation as a closure condition.

We shall consider space-times which are consistent with the exact scaling solutions for the averaged spatial curvature and kinematical backreaction variables as formulated in Ref.~\refcite{morphon2006,larena2009},
\begin{equation}
\label{eq:scalingsol}
\average{\CR}{\CD} \;=\; \laverage{\CR}{\CD} \, a_{\CD}^{n} \ ; \qquad {\CQ}_{\CD} \;=\;{\CQ}_{\initial\CD} \, a_{\CD}^{p}\ , 
\end{equation}
as an ansatz for the needed closure condition, with $n$ and $p$ being real numbers, and $\initial{\CD}$ denoting an initial domain for which the definition $a_{\initial\CD} \equiv 1$ is imposed. 
Plugging the ansatz (\ref{eq:scalingsol}) into the integrability condition (\ref{eq:integrability}) we have that either $n = -\,2$ and $p = -\,6$ or $n = p$ must be satisfied.

The solution $n = -\,2$ and $p = -\,6$ leads to a quasi-Friedmannian model in which the backreaction variable ${\CQ}_{\CD}$ is negligible today (due to its rapid fall-off as a function of volume), and which is the only case where structure formation, encoded in ${\CQ}_{\CD}$, is decoupled from the averaged spatial curvature, such that the quasi-FLRW curvature constraint $(\,\average{\CR}{\CD}a_\CD^2 \,
)^{\bdot} = 0$ is satisfied. 

In the present analysis we will consider the class of solutions $n = p$, which implies coupling of structure formation and averaged scalar curvature. For this class of solutions we have the proportionality relation
\begin{equation}
\label{eq:backreactionpropcurv}
Q_{\CD} \;=\; -\, \frac{n\,+\,2}{n\,+\,6} \,\average{\CR}{\CD} \,
\end{equation}
between kinematical backreaction and averaged spatial curvature.
Thus, positive kinematical backreaction (dominance of the variance in the fluid expansion rate over shear\cite{Buchert2000}) implies negative spatial curvature when $n > -2$. 

It is convenient to introduce the following effective deceleration parameter for characterizing the different possible scaling solutions in terms of their acceleration:\footnote{Parameters evaluated at the present epoch are indexed with $\CD_{\it  0}$ throughout this paper.}
\begin{equation}
\label{eq:decelerationparam}
q_{\CD} \;\equiv\; -\, \frac{{\ddot a}_\CD \, a_\CD }{{\dot a}^2_\CD} \; =\;  \frac{  \Omega_{m}^{\CD}  \,-\, (n+2) \, \Omega_\CX^{\CD} }{2}  \; =\;  \frac{  \Omega_{m}^{\CD_{\it  0}}  \,-\, (n+2) \, \Omega_\CX^{\CD_{\it  0}} \left( \frac{a_\CD}{a_{\CD_{\it  0}}} \right)^{n + 3} }{2 \, \Omega_{m}^{\CD_{\it  0}} \,+\, 2 \, \Omega_\CX^{\CD_{\it  0}} \left( \frac{a_\CD}{a_{\CD_{\it  0}}} \right)^{n + 3}} \ , 
\end{equation}
analogous to the definition of the FLRW deceleration parameter. The second equality follows from combining (\ref{eq:averagedraychaudhuri}) and (\ref{eq:averagedhamilton}), and using the definitions of the cosmological parameters given in Sec. \ref{subsec:cosmologicalparams}. The last equality follows from the scaling conditions (\ref{eq:scalingsol}) with $n = p$, and from (\ref{eq:omX}). 
From (\ref{eq:decelerationparam}), we can formulate the following acceleration condition:
\begin{equation}
\label{eq:accelerationcondition}
q_{\CD} \, < \, 0  \quad \Leftrightarrow \quad    (n+2) \, \left( \frac{a_\CD}{a_{\CD_{\it  0}}} \right)^{n + 3} \, > \, \frac{ \Omega_{m}^{\CD_{\it  0}}}{1 \,-\, \Omega_{m}^{\CD_{\it  0}}}  \ ,
\end{equation}
valid for $0 < \Omega_{m}^{\CD_{\it  0}} <1$. 
Thus, for $n \leq -2$, volume acceleration does not occur at any epoch, as the kinematical backreaction $Q_{\CD}$ is negative in this case. For $n > -2$, acceleration might be reached depending on the value of $\Omega_{m}^{\now\CD}$. 
We note that $n = 0$ results in an acceleration condition formally similar to the flat FLRW model ($\Omega_{\Lambda} = 1 - \Omega_m$) acceleration condition: $2 \, \left( a/ \now a \right)^{3} \, > \,  \Omega_{\now m} / (1 - \Omega_{\now m})$, where $a$ is the FLRW scale factor. This is expected, since the backreaction term $Q_{\CD}$ is constant in this case, and thus acts as an effective cosmological constant in the averaged Raychaudhuri equation (\ref{eq:averagedraychaudhuri}) (\textit{cf.} 
Ref.~\refcite{Buchert2008} [Sec. 3.3.2]).

{We note that the timescape model, which we also investigate in this analysis, is not part of the scaling solutions discussed here (even though it is solution to the set of averaged equations discussed in Sec.~\ref{subsec:buchertsscheme}). Rather it is a two-scale model with volume partitioning into over-dense flat regions and under-dense \say{void regions}. For details about the formulation of the timescape model, see Ref.~\refcite{TS2009forma}.}
\subsection{The template metric \label{ss:metric}} \label{subsec:templatemetric}
In order to translate physical observables of redshift and photon flux into \say{measurements} of the free parameters $n$ and $\Omega_m^\now\CD$ of the scaling solutions outlined in Sec. \ref{subsec:scalingsol}, we must parameterize predictions of the observables in terms of $n$ and $\Omega_m^\now\CD$.  
      
{With knowledge of the entire hierarchy of structure in the Universe and the corresponding inhomogeneous metric, one would in principle be able to do general relativistic ray-tracing, and properly describe the measurements of an observer at a given location without the need for an averaging scheme. In practice we do not have access to such information, and the aim here is to formulate an effective model for light propagation over cosmic scales $\CD \gtrsim 100\,$Mpc/h given knowledge of the functions $a_\CD$, $\average{\varrho}{\CD}$, $\average{\CR}{\CD}$, and ${\CQ}_\CD$ describing the Universe on such scales. 
These global parameters are built from averages of local space-time variables fulfilling the Einstein equations, but are not themselves solutions to any \say{global Einstein equations} valid on the scale $\CD$. Rather, they are solutions to the set of equations (\ref{eq:averagedraychaudhuri}), (\ref{eq:averagedhamilton}), and (\ref{eq:energyconservation}). Thus, $a_\CD$ is \emph{not} to be thought of as a scale factor in a local metric, and $\average{\CR}{\CD}$ is \emph{not} to be thought of as the spatial Ricci curvature built from such a metric. 
We can nevertheless conjecture that light sampling the Universe in a volume averaged sense is, on average, propagating according to null-geodesics of an \emph{effective} metric which reduces to an FLRW $3-$metric described with spatial curvature $\average{\CR}{\CD}$ at each leaf of the space-time normal to the fluid flow, but which has non-trivial union between such leaves due to the non-commutativity of the averaging and time-evolution operations.}

{Based on this conjecture,} we introduce a template metric {for describing light propagation on cosmic scales} as a constant-curvature metric but which, unlike the FLRW solution, allows for curvature evolution in \say{cosmic time}. 
{We stress that the introduction of such a template metric, which is not a solution to Einstein's equations, is not a violation of general relativity. On the contrary, in a general relativistic universe model, any metric theory describing average light propagation on large scales must be effective}.\footnote{{We refer the reader to Ref. \refcite{larena2009} for further motivations for introducing the template metric, where it is discussed how constant-curvature metrics can be obtained via Ricci flow smoothing of Riemannian hypersurfaces\cite{ricciflow}. Even though the template metric described in this section is not solution to Einstein's equations, local metrics of the same form have been studied as solutions to the Einstein equations (see the recent paper by Stichel \cite{stichel} and references therein).}}

The form of the effective metric follows the proposal of Ref. \refcite{larena2009}:
\begin{equation}
\label{eq:metric}
^4g^\CD \; \equiv \; -\, \dd t^2 \,+\, L^2_{H_\now\CD} \, a^2_\CD \, \pa{\, \frac{\dd r_\CD^2}{1 \,-\, \kappa_\CD(t) \, r_\CD^2} \,+\, r_\CD^2 \, \dd\Omega^2 \,}    \,,
\end{equation}
with $t$ being the proper time function of the dust fluid, such that $t =\,$const. selects hypersurfaces orthogonal to the fluid flow, and $r_\CD$ is a dimensionless radial coordinate, which also has the interpretation as a comoving distance; 
$a_{\CD}$ is the dimensionless volume scale factor governed by (\ref{eq:averagedraychaudhuri})--(\ref{eq:integrability}), 
and $a_{\now\CD}  L_{H_{\now\CD}} \equiv H_{\now\CD}^{-1}$ is the present-day Hubble horizon;
$\dd \Omega^2 \,\equiv\,  (\dd \theta^2 + \sin(\theta)^2 \, \dd \phi^2)$ is the angular element on the unit sphere, 
and $\kappa_\CD$ is a dimensionless spatial constant-curvature function 
related to the averaged spatial Ricci scalar through
\begin{equation}
\label{eq:defkappa}
\kappa_\CD (t)\;\equiv\; \,\frac{\average{\CR}{\CD}(t)}{|{\average{\CR}{\now\CD}}|} \, \frac{a^2_\CD(t)}{{a_\now\CD}^2}\ .
\end{equation}
For the class of scaling solutions described in Sec. \ref{subsec:scalingsol}, with $n=p$, one can rewrite $\kappa_\CD$ using (\ref{eq:omX}) and (\ref{eq:backreactionpropcurv}):
\begin{equation}
\label{eq:kappa}
\kappa_{\CD}(a_{\CD}) \;=\;-\,\frac{(n+6) \, \Omega_{\CX}^{\now\CD}}{|(n+6) \, \Omega_{\CX}^{\now\CD}|}\, \frac{a_{\CD}^{(n+2)} }{ a_{\now\CD}^{(n+2)}    }\ .
\end{equation}
In what follows we advance the idealizing conjecture that light propagation over cosmological scales is effectively described by null geodesics in the template metric (\ref{eq:metric}). 
This is an assumption that follows the homogeneous-geometry approximation of the standard model, but corrects for the evolution of curvature to comply with the exact average properties. We also note that more insight and work is needed to improve on this first-step template metric.

\subsection{Distance modulus \label{sub:tabledL}} 
\label{subsec:distancemodulus}
In order to constrain the scaling solutions with supernova data we must make a prediction for the distance modulus within this class of models.  
We will compute the distance modulus as a function of redshift in the template metric of Sec. \ref{subsec:templatemetric}. 

The distance modulus is defined in terms of the luminosity distance $d_L$ in the following way:
\begin{equation}
\label{eq:distmodulus}
\mu(z_\CD) \;  = \; 5 \, \log_{10} \pa{ \frac{d_{L}(z_\CD)} { \SI{10}{\mega \parsec}}  }\,,
\end{equation}
where $z_\CD$ is the redshift as inferred from the domain-dependent scale factor (see the below equation (\ref{eq:zD})).
By Etherington's theorem (see Ref. \refcite{nezihe:lightpropagation} and references therein),
\begin{equation}
\label{eq:dL}
d_L(z_\CD) \;=\; (1 \,+\, z_\CD)^2 \, d_A(z_\CD) \, ,
\end{equation}
where $d_A$ is the angular diameter distance. 
The angular diameter distance is given via the metric (\ref{eq:metric}) as
\begin{equation}
\label{eq:dA}
d_A(z_\CD) \;=\; \frac{1}{H_\now\CD} \, a_\CD(z_\CD) \, r_\CD(z_\CD) \,.
\end{equation}
From the geodesic equation of (\ref{eq:metric}) we have that light emitted and absorbed by observers comoving with the dust, and propagating radially with respect to the central observer, is redshifted by\footnote{We henceforth drop the domain index for the redshift.}
\begin{equation}
\label{eq:zD} 
z_{\CD}(a_{\CD}) \; = \; \frac{\hat{k}^{0}(a_{\CD})}{a_{\CD}} \,-\, 1\; ,
\end{equation}
with $\hat{k}^0$ given by
\begin{equation}
\label{eq:derik}
\frac{\dd \ln(\hat{k}^{0})}{\dd a_{\CD}}\, \; = \; -\, \frac{r_\CD^{2}(a_{\CD})}{2\,(1 \,-\, \kappa_{\CD}(a_{\CD})r_\CD^{2}(a_{\CD}))} \, \frac{\dd \kappa_{\CD}(a_{\CD})}{\dd a_{\CD}} \, .
\end{equation}
The dimensionless coordinate distance $r_\CD$ along the null rays is 
\begin{equation}
\label{eq:dericd}
\frac{\dd r_\CD}{\dd a_{\CD}}\, \;= \; - \, \frac{1}{a_\CD^2} \, \sqrt{\frac{1 \,-\, \kappa_{\CD}(a_{\CD}) \, r_\CD^2(a_{\CD})}{\Omega_{m}^{\now\CD} \, a_{\CD}^{-3} \,+\, \Omega_{\CX}^{\now\CD} \, a_{\CD}^{n}}} \ , \qquad r_\CD \,(a_{\CD} = 1) \, \equiv \, 0  \;,
\end{equation}
where the expression for $\kappa_{\CD}$ (\ref{eq:kappa}) has been used.\footnote{The expression (\ref{eq:dericd}) for $\dd r_\CD / \dd a_{\CD}$ is different from that in Eq. (41) of Ref.~\refcite{larena2009}, which is due to minor typos in Ref.~\refcite{larena2009}; see also the remarks in Ref. \refcite{Chinese2018}. 
}

\section{Methods}
\label{sec:sn}
We use the Joint Light-curve Analysis (JLA) sample\cite{JLA2014} containing $740$ supernovae to test the scaling solutions described in Sec. \ref{sec:formalism}. The JLA catalogue gathers data from four independent studies: SuperNovae Legacy Survey (SNLS), Sloan Digital Sky Survey (SDSS), nearby supernovae (Low--z), and Hubble Space Telescope (HST). 

\subsection{The SALT2 method}
The Spectral Adaptive Lightcurve Template 2 (SALT2) method for making supernovae standard candles consists in fitting the supernovae light-curves to an empirical template, and subsequently using the parameters of the light-curve fit in the empirical model for band correction:
\begin{equation}
\label{eq:distmod_obs}
\mu_{SN} \;=\; m_B^* \,-\, M_B \, + \, \alpha\,x_1 \,-\, \beta\,c \,,
\end{equation}
where $m_B^*$ is the peak of the apparent magnitude in the B-band, $M_{B}$ is the intrinsic magnitude in the rest frame of the supernova, $x_{1}$ is the light-curve stretch parameter, and $c$ is the colour correction parameter for each supernova in the sample.
$m_B^*$, $x_{1}$, and $c$ are obtained from template fitting of the supernovae light-curves\cite{JLA2014}; 
$\alpha$ and $\beta$ are global regression parameters that are determined in the fit. 

\subsection{The Likelihood function \label{sec:code}}
We now briefly review the likelihood function $\CL( \hat{X}  \,|\, \theta)$ used in this analysis, where $\hat{X} = \{\hat{m}_{B,1}^* \, , \hat{x}_{1,1}, \, \hat{c}_1, ... , \hat{m}_{B,N}^* \, , \hat{x}_{1,N}, \, \hat{c}_N\}$ are the \say{observed} parameters for the supernovae labelled $1,...,N$, and $\theta$ is short for all model assumptions (cosmological model, model for band correction, etc.). 

The hats over the parameters in $\hat{X}$ indicate that these parameters are inferred from data, whereas the corresponding parameters without hats represent the \say{true} underlying (or intrinsic) parameters. 

We use the likelihood function as formulated in Ref.~\refcite{Likelihoodconstruction}, with the model for the distribution of intrinsic supernovae parameters proposed in Ref.~\refcite{NGS16_noacceleration}, where the intrinsic parameters $M_B , x_{1} , c$ of each supernova are assumed to be drawn from identical and independent Gaussian distributions with means $M_{0}, x_{1,0}, c_{0}$ and standard deviations $\sigma_{M_0}, \sigma_{x_{1,0}}, \sigma_{c_0}$. Using the SALT2 relation (\ref{eq:distmod_obs}) and the assumptions presented in Ref.~\refcite{NGS16_noacceleration}, the final expression of the likelihood function is
\begin{eqnarray}
\label{eq:likelihoodfinalexpression}
\CL \;=\; &|\, 2 \pi \,(\Sigma_\textnormal{d} \,&+\, A^T  \Sigma_\textnormal{l} A) \,|^{-1/2} \; \nonumber \\ 
&\times \exp &\pac{ -\, (\hat{Z} \,-\, Y_0 A) \, (\Sigma_\textnormal{d} \,+\, A^T \Sigma_\textnormal{l} A)^{-1} \, (\hat{Z} \,-\, Y_0 A)^T/2 \, } \,, \end{eqnarray} 
where $|\,.\,|$ denotes the determinant of a matrix, $\Sigma_\textnormal{d}$ is the estimated experimental covariance matrix (including statistical and systematic errors), and $\Sigma_\textnormal{l}$ is the \say{intrinsic covariance matrix} diag$(\sigma^2_{M_0}, \sigma^2_{x_{1,0}}, \sigma^2_{c_0}, \sigma^2_{M_0}, \sigma^2_{x_{1,0}}, \sigma^2_{c_0}, ...)$ of dimension $3N \times 3N$; 
$\hat{Z} \equiv \{\hat{m}_{B,1}^* \,-\, \mu_1, \hat{x}_{1,1}, \, \hat{c}_1, ... , \hat{m}_{B,N}^* \,-\, \mu_1, \hat{x}_{1,N}, \, \hat{c}_N\}$, where $\mu_1,...,\mu_N$ are the distance moduli evaluated at the measured redshifts $\hat{z}_1,...,\hat{z}_N$ of the supernovae, and $Y_0 \equiv \{M_{0}, x_{1,0}, c_{0},M_{0}, x_{1,0}, c_{0}, ...\}$; $A$ is the blog diagonal matrix
\begin{equation}
\label{eq:defA}
A \;=\; \pa{
\begin{array}{cccc}
1 &  0 & 0 & 0 \\
- \, \alpha & 1 & 0 & 0  \\
\beta & 0 & 1 & 0 \\ 
0 & 0 & 0 & \ddots
\end{array}}\,.
\end{equation}
The final likelihood thus contains the following eight free parameters: $\alpha$, $\beta$, $M_{0}$, $x_{1,0}$, $c_{0}$, $\sigma_{M_0}$, $\sigma_{x_{1,0}}$, and $\sigma_{c_0}$ in addition to the cosmological parameters entering the expression for the distance modulus $\mu$. 

It has been suggested to include even more empirical parameters in order to model redshift-dependence in the intrinsic supernovae parameters and observational biases in these.\cite{RH16_yesacceleration}\footnote{For a discussion on the degeneracy introduced between parameters that describe the supernovae and the cosmological model in such empirical modeling, see Ref.~\refcite{TS2017}.} {In this paper we stick to the likelihood function (\ref{eq:likelihoodfinalexpression}) based on a minimal number of empirical parameters.
We focus on the constraint of cosmological parameters and on the relative quality of fit provided by different cosmological models. For an assessment of the fit of the likelihood function (\ref{eq:likelihoodfinalexpression}) to data, and in particular of the ability to fit the distributions of the measured supernovae parameters $\hat{x}_{1} $ and $\hat{c}$, see Ref.~\refcite{Likelihoodconstruction}.} 

\section{Data analysis} 
\label{sec:modelanalysis}
We now constrain the parameter space of the scaling solutions with the JLA catalogue using the SALT2 relation and the likelihood model specified in Sec. \ref{sec:code}. We then compare the quality of fit to that of the $\Lambda$CDM model, the Milne universe model with no sources and a negative constant curvature (henceforth named the `empty universe model'), and the timescape model. 
We discuss the scales of application of the scaling solutions in relation to the application of a redshift cut in the data in Sec. \ref{subsec:redshiftcut}. In Sec. \ref{subsec:results} we present our results.

\subsection{Statistical homogeneity scale and cut-off in redshift} 
\label{subsec:redshiftcut}
Any model describing light propagation on a given scale should, for the sake of self-consistency, only be applied to 
light-rays propagating over at least that scale. 

Since all the models tested in this analysis have, per construction, structureless geometry and are designed to hold above an approximate statistical homogeneity scale, it is natural (or even mandatory) to impose a cut-off in radius relative to the observer corresponding to the approximate homogeneity scale. Light emitted by supernovae below such a radius is probing scales below which the cosmological averaged metric description applies. 

The largest scales of second-order correlations between structures (applying a cut-off of $\sim 1$\% in the two-point correlation function)\cite{Scrimgeour} is estimated to be $\sim 70-120$ Mpc/h in $\Lambda$CDM.\footnote{Note that higher-order correlations are still significant on Gpc scales. Probed through Minkowski functionals containing all orders of correlation functions, the analysis of SDSS LRG samples revealed more than $2\,\sigma$ deviations from $\Lambda$CDM mock catalogues on scales beyond $600$ Mpc/h \cite{Wiegand}.}
Following Ref.~\refcite{TS2017} we apply a cut-off at a redshift radius in the CMB frame $z_\textnormal{CMB,min} = 0.033$ relative to a central observer, corresponding to a comoving distance of $\sim 100$ Mpc/h{, when computing parameter constraints}.
This choice is a bit more conservative than that imposed in Ref.~\refcite{Riess2007} of $z_\textnormal{CMB,min} = 0.024${, corresponding to $\sim 70$ Mpc/h}. The slight difference in choice of cut-off does not strongly affect the {parameter estimates. We shall examine a few different choices of redshift cut-off when comparing the quality of fit of the tested models, in order to establish the degree of robustness of the results to the subsetting of data.}

\subsection{Results} 
\label{subsec:results}
We use the likelihood function given in Sec. \ref{sec:code} and the equation for the distance modulus (\ref{eq:distmodulus}) to constrain the scaling solutions. 

The $1\, \sigma$ confidence bounds on the cosmological parameters of the scaling solution {(with fixed scaling index $n=-1$ and free scaling index respectively)} are shown in Table \ref{tab:bfpara}, together with the corresponding results for the $\Lambda$CDM model {(with imposed spatial flatness and free curvature parameter $\Omega_k$ respectively)}, the empty universe model, and the timescape model (see Table 2 of Ref.~\refcite{TS2017}). 
It should be noted that the matter cosmological parameters $\Omega_m^\now\CD$ of all the models cannot be directly compared (even though they are represented by the same symbol to ease the notation). The scaling solutions,
the $\Lambda$CDM model, and the timescape
model are non-nested (i.e. none of the models can be obtained from any of the other models by parameter space constraints), and their expansion history depend on $\Omega_m^\now\CD$ in different ways.

{The constrained versions of the scaling solution and the $\Lambda$CDM model, with $n=-1$ and $\Omega_k = 0$ respectively, are associated with much less uncertainty in the $\Omega_m^\now\CD$ parameter than the corresponding unconstrained models. This is due to the coupling of the cosmological parameters in the likelihood function.

In addition to the cosmological parameters, we quote the best-fit \say{nuisance parameters} $\alpha$, $\beta$, $M_{0}$, $x_{1,0}$, $c_{0}$, $\sigma_{M_0}$, $\sigma_{x_{1,0}}$, and $\sigma_{c_0}$, described in Sec.~\ref{sec:code}.
Our best-fit findings are similar to those found in Ref.~\refcite{Likelihoodconstruction,TS2017}, and typical differences between the parameters inferred when assuming the respective cosmological models are within a few percent. For typical $1 \, \sigma$ constraints on the regression coefficients $\alpha$ and $\beta$ of the SALT2 relation (\ref{eq:distmod_obs}) and on the mean $M_{0}$ and width $\sigma_{M_0}$ of the distribution of intrinsic magnitudes, see Ref.~\refcite{JLA2014}. }

\begin{table}[tbh]\centering \footnotesize
\tbl{{Best-fit parameters with a redshift cut-off at $z_\textnormal{CMB,min} = 0.033$. For the cosmological parameters we also quote {\say{1$\,\sigma$}$ = 68.27 ... \%$} confidence bounds. Note that $x_{1, 0}$, $c_{0}$, $M_{B,0}$, $\sigma_{x_{1,0}}$, $\sigma_{c_0}$, and $\sigma_{M_{B,0}}$ are the mean and width parameters of the assumed Gaussian distributions from which the intrinsic parameters of each supernova are assumed to be drawn (see Sec.~\ref{sec:code}). Thus, the numbers quoted for $\sigma_{x_{1,0}}$, $\sigma_{c_0}$, and $\sigma_{M_{B,0}}$ are best-fit values of the widths of these Gaussian distributions and \emph{not} error bars on the best-fit determinations of $x_{1, 0}$, $c_{0}$, and $M_{B,0}$.\hfill \hfill} }{
\begin{tabular}{ccccccc}
\noalign{\vspace{0.2cm}} \hline \noalign{\vspace{0.2cm}}
\multirow{1}{*}{Models} 						& Scaling solution 					& {Scaling solution}  			& $\Lambda\text{CDM}$ 			& $\Lambda$CDM 									& Empty  			& Timescape \\
											&  									& {$n\,=\,-\,1$} 				& 								& ${\Omega_k\, =\, 0}$ 							& Universe 			& 			\\
\noalign{\vspace{0.2cm}} \hline \noalign{\vspace{0.2cm}} 
$\Omega_m^\now\CD$ or $\Omega_{\now m}$  	& {$0.24^{+0.13 \, }_{-0.24 \, }$}	& {$0.25^{+0.04}_{-0.04 }$} & ${0.30^{+0.10}_{-0.11}}$	& $0.37^{+0.03}_{-0.03}$								& -				& $0.31^{+0.07 \, }_{-0.09 \, }$ 	\\\noalign{\vspace{0.1cm}}
$n$ 										& {$-\,1.0^{+0.7 \, }_{-0.6 \, }$}	& -						& - 								& - 			 										& - 					& - 								\\\noalign{\vspace{0.1cm}}
$\Omega_k$ 								& - 									& {-} 						& ${0.17^{+0.28}_{-0.26}}$ 	& - 			 										& - 					& - 								\\\noalign{\vspace{0.1cm}}
$\alpha$ 									& $0.13$ 							&  ${0.13}$					& ${0.13}$					& $0.13$ 											& $0.13$ 			& $0.13$ 						\\\noalign{\vspace{0.1cm}}
$x_{1,0}$ 									& $0.11$ 							& ${0.11}$					& ${0.11}$					& $0.11$	 										& $0.10$ 			& $0.11$ 						\\\noalign{\vspace{0.1cm}}
$\sigma_{x_{1,0}}$ 							& $0.90$ 							& ${0.90}$					& ${0.90}$					& $0.90$ 											& $0.90$ 			& $0.90$ 						\\\noalign{\vspace{0.1cm}}
$\beta$ 										& $3.1$ 								& ${3.1}$					& ${3.1}$					& $3.1$ 												& $3.1$ 				& $3.1$ 							\\\noalign{\vspace{0.1cm}}
$c_0$ 										& $-\,0.021$ 						& ${-\,0.021}$				& ${-\,0.021}$				& $-\,0.022$ 										& $-\,0.020$ 		& $-\,0.021$ 					\\\noalign{\vspace{0.1cm}}
$\sigma_{c_0}$ 								& $0.069$ 							& ${0.069}$					& ${0.068}$					& $0.069$ 											& $0.069$ 			& $0.069$ 						\\\noalign{\vspace{0.1cm}}
$M_{B,0}$ 									& $-\,19$ 							& ${-\,19}$					& ${-\,19}$					& $-\,19$  											& $-\,19$ 			& $-\,19$ 						\\\noalign{\vspace{0.1cm}}
$\sigma_{M_{B,0}}$  							& $0.10$ 							& ${0.10}$					& ${0.10}$					& $0.10$ 											& $0.11$ 			& $0.10$ 						\\\noalign{\vspace{0.2cm}} \hline
\end{tabular}}
\label{tab:bfpara}
\end{table}

\begin{figure}[ht]
\centering
\includegraphics[width=0.9\textwidth]{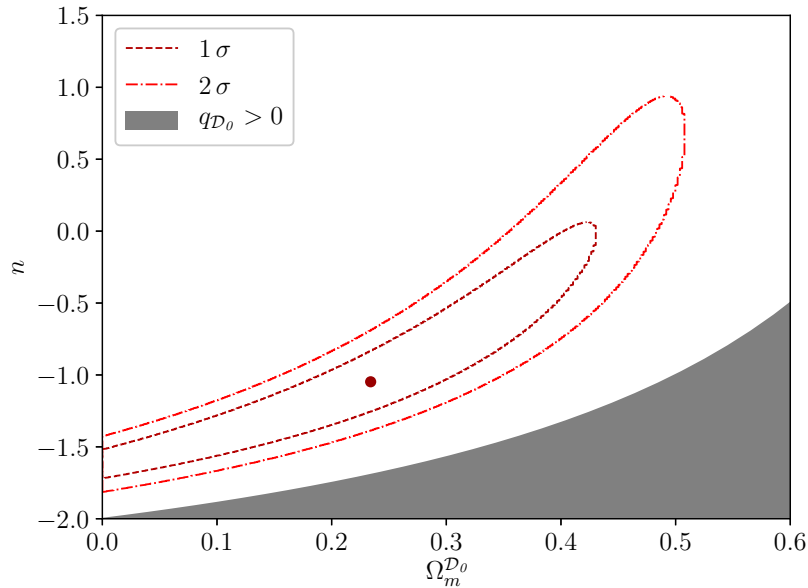}
\caption{$1 \, \sigma$ and $2 \, \sigma$ confidence contours of the parameters $\Omega_m^{\now\CD}$ and $n$ of the scaling solutions. The best-fit, $\{\Omega_m^\now\CD = 0.24 \, ,\,n=-1.0\}$, is marked by a dot. The shaded area represents models with present-epoch volume deceleration $q_{\CD_{\it  0}} > 0$, and the remaining area of the parameter space has positive present-epoch volume acceleration. Thus, deceleration is ruled out at the $> 2 \, \sigma$ level for the class of scaling solutions tested.}
\label{fig:diagOmn}
\end{figure}

The frequentist $1 \, \sigma$ and $2 \, \sigma$ confidence contours for the scaling solutions are shown in Fig. \ref{fig:diagOmn}. Our results are consistent with positive present-epoch volume acceleration, ruling out deceleration at the $> 2 \, \sigma$ level, for the class of scaling solutions tested.

A striking result is that the best-fit scaling index {$n = -\,1.0^{+0.7 \, (1\sigma)}_{-0.6 \, (1\sigma)}$} is consistent with the results obtained in Ref.~\refcite{Lagrangian2013} in a perturbative framework around an Einstein-de Sitter background\footnote{{Einstein-de Sitter (flat \say{matter only} FLRW model with $\Omega_m = 1$ and $\Omega_{\Lambda} = 0$) exhibits volume deceleration, and constitutes an interesting background model for studying the possible emergence of spatial curvature and volume acceleration from an initially decelerating and (almost) spatially flat universe model.}}, where the leading-order (or largest-scale) backreaction was found to obey the scaling law $\CQ_\CD \;\propto\;a_{\rm EdS}^{-1}$. The best-fit scaling index is thus compatible with a perturbative evaluation of backreaction (extrapolating  the perturbative scaling law). {Notice also that the best-fit scaling index is consistent with $n=0$ at the $2 \, \sigma$ level (but not at the $1 \, \sigma$ level). For this value of $n$, backreaction is mimicking a cosmological constant, \textit{cf.} Ref. \refcite{Buchert2008} [Sect. 3.3.2].}

Comparing Fig. \ref{fig:diagOmn} with the contour plot of Ref.~\refcite{larena2009} {showing the constraints of the scaling solutions from} CMB data from WMAP3-yr data and $71$ SNIa from the SNLS Collaboration, there is a significant amount of overlap of the $2 \, \sigma$ contours. However, the volume of the likelihood in the present analysis is shifted towards lower values of $\Omega_m^{\now\CD}$ and $n$ as compared to Ref.~\refcite{larena2009}.\footnote{It is difficult to compare with the results of Ref.~\refcite{larena2009} because of the sparse supernova sample used and since the best-fit is obtained from a combination of the supernova and CMB data. Moreover, the error-bars on the best-fit parameters are not quoted. Generally, we refer to Ref.~\refcite{larena2009} with respect to the theoretical foundations and methods, the results obtained therein are by now outdated.} 

{Comparing the results of Table \ref{tab:bfpara} with the constraints on the scaling solutions from Ref.~\refcite{Chinese2018}, obtained from measurements of the Hubble parameter from the differential age method and radial baryon acoustic oscillation data, we find agreement within $1 \, \sigma$ of both the scaling index $n$ and the matter cosmological parameter $\Omega_m^{\now\CD}$.}

We compare the quality of fit of the scaling solutions with that of the $\Lambda$CDM model, the empty universe model, and the timescape model using the \textit{Akaike Information Criterion}\footnote{{The Akaike Information Criterion is one of many methods valid for model selection. For an overview of some common methods and their interpretations see Ref.~\refcite{KerscherWeller}. } } (AIC)\cite{Akaike}.
The AIC is a measure of the relative probability of minimal information loss between two models:
\begin{equation}
\label{AICratio_prob}
\frac{p_1}{p_2} \;=\; \frac{ \exp(-\, \textnormal{AIC}_1 \,/\, 2)} {\exp( -\, \textnormal{AIC}_2 \,/\, 2)}\ ; \qquad \textnormal{AIC}_i \;\equiv\; 2 \, q_i \,-\, 2 \, \ln( \CL_i)\ ,
\end{equation}
where $q_i$ is the number of parameters and $\CL_i$ is the maximum likelihood of model $i$, where $p_i$ is the probability that model $i$ minimizes the (estimated) information loss, and where the two models are labelled $i=1,2$, respectively.
The AIC relative likelihood measure (\ref{AICratio_prob}) can be viewed as a generalization of the likelihood ratio to 
non-nested models. 

The interpretation of the relative numerical estimates of the AIC measure for different models is context-dependent. As a rough guideline, differences in AIC between two models of at least $2$, $6$, and $10$ (corresponding to the AIC relative likelihood {with the most likely model in the denominator} not exceeding $0.4$, $0.05$, and $0.007$, respectively) are characterized as providing \say{positive}, \say{strong}, and \say{very strong} evidence, respectively, in favour of the model with minimal AIC\cite{KassRaftery}.\footnote{{When the AIC likelihood is bigger than one -- i.e. the most likely model is in the numerator -- the interpretation reverses such that models with AIC relative likelihood \emph{not smaller} than $1/0.4$, $1/0.05$, and $1/0.007$ are characterized as providing \say{positive}, \say{strong}, and \say{very strong} evidence, respectively, in favour of the model with minimal AIC.}}

We show the results of the AIC values and the AIC relative likelihood measure in Table \ref{tab:AIC}. We use {both} the spatially flat $\Lambda$CDM model {and the $\Lambda$CDM model with free curvature parameter $\Omega_k$ as references, and quote $p_\textnormal{model} / p_\textnormal{$\Lambda$CDM}^{\Omega_k =\, 0}$ and $p_\textnormal{model} / p_\textnormal{$\Lambda$CDM}$ for each of the models.
The results are shown for data excluded below redshift cuts, $z_\textnormal{CMB,min}$, of $0.024$, $0.033$, $0.07$, and $0.15$, respectively, to examine the robustness of the AIC results to different redshift cuts in data. 
The values of $z_\textnormal{CMB,min}$ of $0.024$ and $0.033$ are two different estimates of the statistical homogeneity scale as discussed in Sec. \ref{subsec:redshiftcut}. 
$z_\textnormal{CMB,min} = 0.07$ and $z_\textnormal{CMB,min} = 0.15$ correspond to $\sim 200$ Mpc/h and $\sim 500$ Mpc/h, respectively, in the concordance $\Lambda$CDM model. These scales might be motivated as conservative homogeneity scale estimates based on the studies of convergence of bulk flow \cite{Kashlinsky2009} and of higher order correlation functions \cite{Wiegand}.}

In addition to the scaling solution with two free cosmological parameters, $n$ and $\Omega_m^{\now\CD}$, the AIC is also computed for the nested solution within this class with the scaling index $n$ fixed to its large-scale theoretical expectation, $n=-1$, from Lagrangian perturbation theory studies.\footnote{ {We refer here to Lagrangian perturbations on an Einstein-de Sitter background investigated in Ref. \refcite{Lagrangian2013}, as discussed above in this section, where $n=-1$ was found to describe the large-scale behaviour of kinematical backreaction and averaged scalar curvature. In this study, the backreaction functionals were derived using the averaged Einstein equations without restricting assumptions together with a closure condition for the averaged system in terms of a first-order Lagrangian perturbation scheme as a realistic model for structure formation. }}

\begin{table}[tbh]
\small
\centering
\tbl{Number of parameters, AIC value, and the AIC relative likelihood {for the cosmological models tested, quoted for four different redshift cuts of data}. The AIC relative likelihood is shown with the spatially flat $\Lambda$CDM model as reference {and with the $\Lambda$CDM model with free curvature parameter $\Omega_k$ as reference respectively.
For each redshift cut, the corresponding approximate $\Lambda$CDM comoving distance to that redshift is shown in parenthesis. The number of supernovae left in the sample after each redshift cut is also shown.}  \hfill \hfill \label{tab:AIC}}
{\begin{tabular}{m{2.4cm}cccccc}
\noalign{\vspace{0.2cm}} \hline \noalign{\vspace{0.2cm}}
\multirow{1}{*}{Models} & Scaling solution & Scaling solution  & $\Lambda\text{CDM}$ & $\Lambda$CDM & Empty  & Timescape \\
&  & $n\,=\,-\,1$ & & ${\Omega_k\, =\, 0}$ & Universe & \\
\noalign{\vspace{0.2cm}} \hline \noalign{\vspace{0.2cm}}
Number of parameters & $10$ & $9$ & ${10}$ & ${9}$ & $8$ & $9$ \\
\noalign{\vspace{0.2cm}} \hline \noalign{\vspace{0.2cm}} \multicolumn{7}{c}{{Redshift cut: $0.024 \; (\sim 70 \text{ Mpc/h}) \quad$ - $\quad 687$ SNIa}} \\ \noalign{\vspace{0.2cm}}
{AIC}   			& ${ -\,213} $					& ${-\,215} $ 				& ${-\,214} $ 						& $ {-\,216} $ 			& ${-\,217} $ 			& ${-\,215}$ \\ 
\noalign{\vspace{0.2cm}}
${p_\textnormal{model} / p_\textnormal{$\Lambda$CDM}^{\Omega_k =\, 0}}$	& ${0.5} $ & ${1.3} $ & ${0.6} $ & ${1.0} $ & ${0.1} $ & ${1.3} $ \\ 
\noalign{\vspace{0.1cm}}
${p_\textnormal{model} / p_\textnormal{$\Lambda$CDM}}$ 					& ${0.8} $ & ${2.3} $ & ${1.0} $ & ${1.8} $ & ${0.2} $ & ${2.4} $  \\
\noalign{\vspace{0.2cm}} \hline \noalign{\vspace{0.2cm}} \multicolumn{7}{c}{{Redshift cut: $0.033 \; (\sim 100 \text{ Mpc/h}) \quad$ - $\quad 655$ SNIa}} \\  \noalign{\vspace{0.2cm}}
{AIC}    			& $ -\,225 $ 							& $ -\,227 $ 						& $ {-\,226} $						& $ -\,228 $ 					& $ -\,229 $ 				& $ -\,227 $				\\ 
\noalign{\vspace{0.2cm}}
${p_\textnormal{model} / p_\textnormal{$\Lambda$CDM}^{\Omega_k =\, 0}}$	& $0.4 $ & $1.0 $ & ${0.5} $ & $1.0 $ & $0.1 $ & $1.0 $ \\ 
\noalign{\vspace{0.1cm}}
${p_\textnormal{model} / p_\textnormal{$\Lambda$CDM}}$ 					& ${0.8} $ & ${2.1} $ & ${1.0} $ & ${2.2} $ & ${0.2} $ & ${2.1} $\\
\noalign{\vspace{0.2cm}} \hline \noalign{\vspace{0.2cm}} \multicolumn{7}{c}{{Redshift cut: $0.07 \; (\sim 200 \text{ Mpc/h}) \quad$ - $\quad 613$ SNIa}} \\  \noalign{\vspace{0.2cm}}
{AIC}   			& $ {-\,233}$ 					& ${ -\,235}$ 				& ${-\,233}$ 							& $ {-\,235} $ 			& ${-\,236}$ 			& $ {-\,235}$ 			\\ 
\noalign{\vspace{0.2cm}}
${p_\textnormal{model} / p_\textnormal{$\Lambda$CDM}^{\Omega_k =\, 0}}$	& ${0.5}$ & ${1.4}$ & ${0.6}$ & ${1.0}$ & ${0.4}$ & ${1.5}$ \\ 
\noalign{\vspace{0.1cm}}
${p_\textnormal{model} / p_\textnormal{$\Lambda$CDM}}$ 					& ${0.9}$ & ${2.5}$ & ${1.0}$ & ${1.8}$ & ${0.7}$ & ${2.6}$ \\
\noalign{\vspace{0.2cm}} \hline \noalign{\vspace{0.2cm}} \multicolumn{7}{c}{{Redshift cut: $0.15 \; (\sim 500 \text{ Mpc/h}) \quad$ - $\quad 514$ SNIa}} \\ \noalign{\vspace{0.2cm}}
{AIC}   			& $ {-\,197} $ 					& ${ -\,199} $ 				& ${ -\,197} $ 						& $ {-\,199} $ 			& ${ -\,195} $ 		& ${ -\,199} $ \\ 
\noalign{\vspace{0.2cm}}
${p_\textnormal{model} / p_\textnormal{$\Lambda$CDM}^{\Omega_k =\, 0}}$	& ${0.3} $ & ${0.8} $ & ${0.4} $ & ${1.0} $ & ${0.1} $ & ${0.7} $ \\ 
\noalign{\vspace{0.1cm}}
${p_\textnormal{model} / p_\textnormal{$\Lambda$CDM}}$ 					& ${0.8} $ & ${2.1} $ & ${1.0} $ & ${2.7} $ & ${0.3} $ & ${1.8} $  \\
\noalign{\vspace{0.2cm}} \hline 
\end{tabular}
 }
\end{table}

{From Table \ref{tab:AIC} we see that all tested models are relatively close in AIC probability. No model has \say{strong} evidence (i.e. AIC relative likelihood of $\leq 0.05$) over another from an AIC perspective for any given redshift cut. 
Some models are weakly preferred over others. For example, the spatially flat $\Lambda$CDM model, the scaling solution with $n=-1$ and the timescape model are all weakly preferred (AIC relative likelihood of $\leq 0.4$) over the empty universe model. 

For the values $z_\textnormal{CMB,min} = 0.033$ and $z_\textnormal{CMB,min} = 0.15$, the spatially flat $\Lambda$CDM is weakly preferred (AIC relative likelihood of $\leq 0.4$) over the scaling solution. However, this conclusion is not robust to the choice of redshift cut, as can be seen in Table \ref{tab:AIC}. Furthermore, these weak preferences vanish when we instead compare the scaling solution to the $\Lambda$CDM model with curvature, which is perhaps the more natural choice, since the models compared in this case have the same number of free parameters and a \say{curvature} parameter each ($n$ and $\Omega_k$ respectively).

The AIC relative likelihoods are in general smaller when quoted with the spatially flat $\Lambda$CDM model as reference than for the $\Lambda$CDM model with curvature as reference, since the likelihood does not increase sufficiently in $\Lambda$CDM by adding the curvature parameter to account for the AIC punishment factor for adding an additional parameter. We note, however, that the best-fit $\Lambda$CDM model has negative curvature (see Table \ref{tab:bfpara}), which is also a feature of the scaling solution.

We conclude that the $\Lambda$CDM model, the scaling solution, and the timescape model provide adequate fits to data.
The spatially flat $\Lambda$CDM model, the scaling solution with $n=-1$ and the timescape model overall have the highest AIC likelihoods. 
The empty universe model is mildly disfavoured as compared to the other models. 

It is important to point out, that the comments made here on the quality of fit are valid for the luminosity distance probed at $z \lesssim 1$ only. For example, the empty universe model is not viable as a cosmological model (for physical reasons and from a quality of fit perspective) for describing CMB physics.

Our findings align with the conclusions in the recent investigation {of Ref.~\refcite{PantheonHuiller} in which it is found that the Pantheon sample probing the $z \lesssim 1$ range is little constraining}, allowing for possibly large 
deviations from $\Lambda$CDM, and with the results of Ref.~\refcite{NGS16_noacceleration} reporting marginal evidence for acceleration {from supernovae alone} within the FLRW framework.

We emphasize that neither the scaling solutions, the timescape model, nor the empty universe model have any local energy-momentum component violating the strong energy condition.

\subsection{Discussion}
Further studies are needed in order to assess the quality of fit of the scaling solutions to a broader range of cosmological data probing different regimes of the expansion history. 

A comment is in order in relation to the combined analysis of cosmological data probing a hierarchy of scales for models that include backreaction. 
Within the standard model it is relatively straightforward to constrain the \say{background} FLRW model with data on various scales: by assumption, the Universe --- apart from in the immediate vicinity of compact objects with $GM/(rc^2) \gtrsim 1$, where $M$ is the mass of the object, and $r$ is its proper physical radius --- is described by a single FLRW background solution with Newtonian potentials, even in the regime where typical density contrasts are highly non-linear.  

In inhomogeneous cosmology, such assumptions are not made. Rather, it is considered a possibility that non-linear structure, through its coupling to the inhomogeneous geometry, 
can significantly impact the appropriate averaged model for describing collective dynamics of a given space-time domain.\footnote{{Note that the hierarchical structure of our space-time can lead to non-trivial general relativistic phenomena, even though each level of the hierarchy is well described as a \say{weak field} perturbation of the previous level \cite{Korzynski_embedding}. Note also, that even though regions containing general relativistic compact sources are negligible in terms of volume measure as compared to the total volume of a given spatial domain, the domain can exhibit non-trivial general relativistic behaviour \cite{Korzynski_compact}.}}

Because of the failure of one simple global metric to serve as a \say{background} cosmological solution for all structure with $GM/(rc^2) \ll 1$ in inhomogeneous cosmology, a coherent scaling solution framework for interpreting physics on a hierarchy of scales is not within immediate reach.
For space-times with a notion of \emph{statistical} homogeneity and isotropy, we might nevertheless expect convergence of the scaling solutions on the largest scales, such that the cosmological parameters (\ref{eq:omega}) and the associated template metric (\ref{eq:metric}) are effectively independent of the spatial scale $\CD$ above an appropriate cutoff in physical size of the domain $\CD$. 
Thus, the scaling solution being valid on the largest scales might be constrained with complementary cosmological data such as  supernovae, galaxy surveys, and CMB data, as long as the given survey probes large enough spatial domains. 

As is well-known, joint fits of the FLRW model with perturbations face the problem of `tensions', e.g. with respect to different values of the Hubble parameter, a problem that we trace back to naive extrapolation of the model from high to low redshifts and from large to small scales. In particular, insufficient modelling of differential expansion of space might be the cause of the `tension' related to the Hubble parameter, \textit{cf.} Refs.~\refcite{tensions,Lagrangian2013,bolejko,bolejkojan,kasaitoshi}.

}

\section{Testing curvature dynamics with upcoming surveys} 
\label{sec:curvaturedynamics}
It is of observational interest to investigate possible signatures distinguishing between models with dynamical spatial curvature and FLRW models (with rigid spatial curvature). 

To test the FLRW constant spatial curvature hypothesis, we can {consider the useful curvature statistic}:\cite{Clarkson}
\begin{equation}
\label{eq:k}
k_H \;=\; \frac{1}{D^2}\left(1 - \left(\frac{\dd D}{\dd z}\frac{H}{\now H}\right)^2 \right)\  , 
\end{equation}
where $D$ is the dimensionless FLRW comoving {transverse} distance related to the angular diameter distance $d_A$ by $D = \now H/c \, (1+z)d_A$, and where $H$ is the FLRW Hubble parameter. {From the expression for the FLRW comoving distance $D = 1/ \sqrt{\Omega_{\now k}} \, \sinh (  \sqrt{\Omega_{\now k}} \, \int_{0}^{z} dz' \frac{\now H}{H(z')} )$ it follows that} $k_H = - \Omega_{\now  k}$ per construction. 
{Note that the equality $k_H = - \Omega_{\now  k}$ is based purely on geometrical identities valid for the FLRW class of models, and thus does not depend on details of the matter content, dark matter equation of state, or other tuneable features within FLRW cosmology.}

For any other model with a prediction of angular diameter distance and volume expansion as a function of redshift, we might also construct the function $k_H$ (\ref{eq:k}). In general, $k_H$ is not interpreted as a spatial curvature density parameter, but simply as the combination of distance measures given by the right-hand side of (\ref{eq:k}), and it is in principle allowed to vary arbitrarily with redshift. 

The function $k_H(z)$ is derivable from $H(z)$ and $D(z)$, and thus, it contains no new information with respect to these two functions. However, $k_H(z)$ is a particularly powerful combination of distance measures, as a 
$k_H(z) \neq \,$const. detection would be a \say{smoking gun} for FLRW geometry violation.

\begin{figure}[!htb]
\begin{subfigure}[b]{\textwidth} 
\includegraphics[scale=0.55]{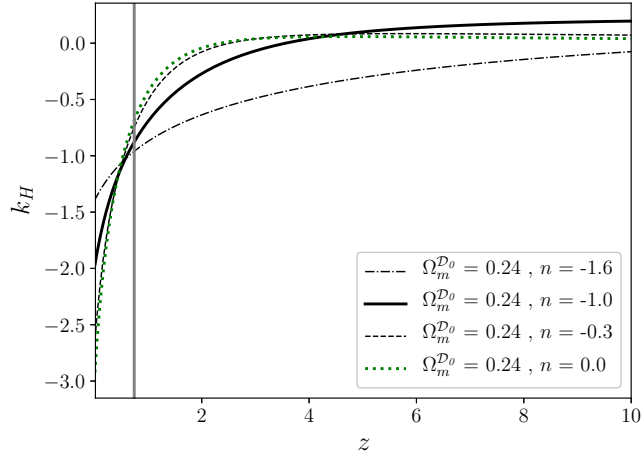}\centering
\caption{The function $k_H$, equation (\ref{eq:k}), predicted by the best-fit scaling solution found in this paper. For a FLRW model universe $k_H = - \Omega_{\now k}$, where $\Omega_{\now k}$ is the spatial curvature density parameter evaluated at the present epoch.}
\label{fig:kevolution}
\end{subfigure}
\medskip
\begin{subfigure}[b]{\textwidth} 
\includegraphics[scale=0.55]{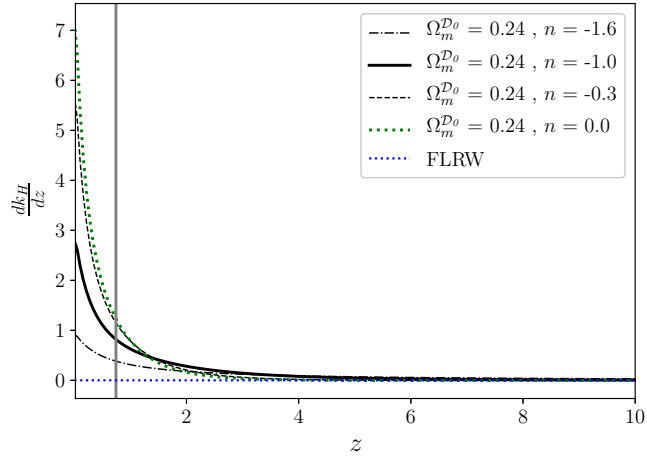}\centering
\caption{$\dd k_H / \dd  z$ as predicted by the best-fit scaling solution found in this paper. In a FLRW model universe 
$\dd  k_H / \dd z = - \dd  \Omega_{\now k} / \dd  z = 0$.}
\label{fig:kprime}
\end{subfigure}
\caption{$k_H$ and $\dd  k_H / \dd  z$ as a function of redshift for the best-fit scaling solution, $\{\Omega_m^\now\CD = 0.24 , n=-1.0\}$. 
The scaling index upper and lower {$1 \, \sigma$} confidence bounds, {$n = - 1.0^{+0.7}_{-0.6}$}, are shown for fixed $\Omega_m^\now\CD$. The solution for $\{\Omega_m^\now\CD = 0.24 ,n=0\}$ is shown as well as a reference.
The vertical grey line marks the redshift of transition from volume deceleration to volume acceleration as predicted by the best-fit model. 
}
\label{fig:kHtest}
\end{figure}

Computing $r_{\CD}$ and $H_{\CD}$ for the best-fit scaling solution, $\{\Omega_m^\now\CD = 0.24 \, ,\,n=-1.0\}$, and substituting $D= H_{\CD_{\it  0}}/c \, (1+z)d_A =H_{\CD_{\it  0}}/c \, \hat{k}^{0} \, r_{\CD}$ and $H = H_{\CD}$ in (\ref{eq:k}), we obtain $k_H$ and $\dd  k_H / \dd  z$ as a function of redshift as shown in Fig.~\ref{fig:kHtest}. 
We also show the {$1 \, \sigma$} confidence bounds on $n$ while keeping $\Omega_m^\now\CD$ fixed. ({The functions  $k_H$ and $\dd  k_H / \dd z$ are relatively robust to varying $\Omega_m^\now\CD$ within its {$1 \, \sigma$} confidence bounds}.)
Note that the JLA sample contains supernovae at redshifts $z\lesssim 1.3$. We nevertheless show the prediction of $k_H$ for higher redshifts.

The evolution of $k_H$ of the best-fit scaling solution is far from the constant-$k_H$ signature of an FLRW model.
The effective curvature parameter $k_H$ tends to increasingly negative values when approaching the present epoch $z \rightarrow 0$, and tends to a constant close to zero in the early universe limit. 

The deceleration parameter (\ref{eq:decelerationparam}) decreases with decreasing redshift and becomes negative at $z\sim 0.7$ for the best-fit scaling solution, $\{\Omega_m^\now\CD = 0.24 \, ,\,n=-1.0\}$, marking the transition between volume deceleration to volume acceleration in the best-fit model. This redshift of transition is comparable to that predicted by {the best-fit} $\Lambda$CDM {model}. 

Interestingly, our results for the scaling solutions show tendencies similar to those of Ref.~\refcite{MontanariRasanen} (see their Fig.~6) where model-independent fitting functions are used to determine the best-fit shape of $k_H$ from the JLA sample, SDSS-III BOSS BAO measurements, and differential age measurements of galaxies. 
In the model-independent determination of $k_H$ in Ref.~\refcite{MontanariRasanen}, negative values of $k_H$ are favoured  towards lower redshifts as shown in their Fig.~6, consistent with our Fig.~\ref{fig:kevolution}.  
Despite these best-fit tendencies in Ref.~\refcite{MontanariRasanen}, the $\Lambda$CDM $k_H = 0$ curvature constraint is still satisfied within the {$2 \, \sigma$} confidence intervals of their analysis using present data.

We emphasize that the best-fit scaling index $n=-1.0$ is obtained when assuming the model to be a single-scaling solution.
More refined modeling of inhomogeneities, e.g. in terms of two-scale volume partitioning into overdense and underdense regions \cite{WiegandBuchert,Krastanov}, feature an additional effect due to the expansion variance between the two regions that adds volume acceleration. {This feature tends to push the best-fit overall scaling index to values closer to $0$, at which backreaction acts as a cosmological constant in the averaged Raychaudhuri equation (\ref{eq:averagedraychaudhuri}).}

Although the distance
modulus--redshift relation of multi-scale models is different, and thus these refined models cannot be directly compared with the single-scaling solution, we show the reference line $n=0$ in Fig.~\ref{fig:kHtest} to illustrate that {the $n=0$ solution closely resembles the solution upper limit of the $1 \, \sigma$ confidence interval $n = -0.3$ found in this analysis.}

The transition from zero FLRW curvature signature $k_H\sim 0$ to negative FLRW curvature signature $k_H \lesssim -1$ {becomes sharper when n tends to zero}; it may therefore be easier to observationally distinguish this case from the constant $k_H$ signature of a FLRW model.

One might estimate $k_H(z)$ cosmology-independently by fitting an empirical function, such as a polynomial truncated at some order, with sufficient freedom for luminosity-distance measurements (from e.g. supernova light-curves) and expansion rate measurements (from e.g. BAO analysis and differential age data), respectively, as done in Ref.~\refcite{MontanariRasanen}. It is especially important for this consistency test that the distance and expansion measurements are indeed cosmology-independent and do not rely on fiducial FLRW assumptions, as the procedure might otherwise circularly confirm the FLRW consistency relation.

With next generation data (such as upcoming surveys from LSST and Euclid \footnote{See Ref.~\refcite{EuclidTheoryWG} for performance forecasts for the Euclid satellite and for a discussion of testable alternative frameworks, hereunder backreaction models, to that of the $\Lambda$CDM model.}) the predictions of Fig.~\ref{fig:kHtest} and complementary distance combinations will be useful for discriminating between the $\Lambda$CDM model, the scaling solutions, as well as other models with non-trivial curvature evolution. 

\section{Conclusion}  
\label{sec:conclusion}
We have investigated the fit of the scaling solutions, which are a class of solutions for the evolution of averaged cosmological variables, constrained by the exact average properties of Einstein's equations and supplemented with a compatible but idealized template metric, to the Joint Light-curve Analysis (JLA) sample of $740$ SNIa. 

We find constraints that are in good agreement with previously found constraints {for the scaling solutions} based on SNIa, CMB, the differential age method, and baryon acoustic oscillation measurements in Ref.~\refcite{larena2009,Chinese2018}.
{Thus, the scaling solutions provide a self-consistent fit to current and complementary cosmological data.}

Our result for the scaling index $n$ is consistent with theoretical expectations on the large-scale behaviour of backreaction within an averaged Lagrangian perturbation approach, Ref.~\refcite{Lagrangian2013}.

Comparing the {quality of} fit of the scaling solutions, the $\Lambda$CDM model and the timescape model, {we find no significant preference of one model over the other from an Akaike Information Criterion (AIC) perspective.
The empty universe model is mildly disfavoured when compared to the fit of the other models.}
This suggests that a broad variety of models of the recent epoch expansion history can match currently available supernova data. {More work is needed in order to assess the quality of fit of the scaling solutions relative to that of $\Lambda$CDM for complementary cosmological data to that of supernovae.}

Backreaction models, exemplified by scaling solutions that match JLA data, predict a clear signature in terms of a particular FLRW curvature consistency measure if compared with the FLRW class of space-times. This indicates that one might be able to significantly discriminate between models with evolving curvature and models with 
constant-curvature geometry with upcoming surveys using this measure.

\vspace{10pt}
{\bf Acknowledgements:}
This work is part of a project that has received funding from the European Research Council (ERC) under the European Union's Horizon 2020 research and innovation programme (grant agreement ERC advanced grant 740021--ARTHUS, PI: TB). CD would like to thank all colleagues at CRAL--ENS for the nice ambiance at work during the summer internship.
AH is supported by an University of Canterbury doctoral scholarship, and acknowledges hospitality for visits to CRAL--ENS, Lyon, supported by Catalyst grant CSG--UOC1603 administered by the Royal Society of New Zealand. 
AH is grateful for the support given by the funds: `Torben og Alice Frimodts Fond', `Knud H{\o}jgaards Fond', and `Max N{\o}rgaard og Hustru Magda N{\o}rgaards Fond'.
We would like to thank the anonymous referee for numerous constructive remarks and suggestions.


\bibliographystyle{ws-ijmpd}

\end{document}